\documentclass[11pt]{article}

\begin{document}


\setlength{\baselineskip}{15pt}

\title{Edge solitons in the QHE\footnote{Talk given by P. Horv\'athy at the meeting
NONLINEARITY, INTEGRABILITY AND ALL THAT \dots.
Gallipoli (Lecce) Italy, 1999.
Ed. M. Boiti, L. Martina, F. Pempinelli, B. Prinari \& G. Soliani,
p. 474.
Singapore ~: World Scientific (2000).}}

\author{
M.~Hassa\"\i ne, P.~A.~Horv\'athy 
 and
J.-C.~Y\'era
\\
\\
Laboratoire de Math\'ematiques et de Physique Th\'eorique
\\
Universit\'e de Tours\\
Parc de Grandmont,
F--37200 TOURS (France)\\
e-mail: horvathy@univ-tours.fr
\\
}


\maketitle

\begin{abstract}
The spacelike reduction of the Chern-Simons Lagrangian 
yields a modified Nonlinear Schr\"odinger Equation (jNLS) where
in the non-linearity the particle density is replaced by the current.
When the phase is linear in the position, this latter is an ordinary NLS
with  time-dependent coefficients which admits interesting solutions,
whose arisal is explained by the conformal properties of 
non-relativistic spacetime. Only the usual travelling soliton
is consistent with the jNLS but adding a six-order potential converts 
it into an integrable equation.

\end{abstract}


\section{Reduction of Chern-Simons}

The Landau-Ginzburg theory of the Quantum Hall Effect \cite{ZHK}
uses the Chern-Simons Lagrangian in $(2+1)$ dimensions,
\begin{equation}
{\cal L}=\frac{1}{4\kappa}\epsilon_{\mu\nu\rho}F_{\mu\nu}A_{\rho}
+
i\phi^\star D_t\phi-{1\over2}\vert\vec{D}\phi\vert^2
-V(\phi), 
\end{equation}
where the scalar field $\phi$ is the order parameter and $A_\mu$
  is the statistical gauge field;
$D_\mu=\partial_\mu-iA_\mu$ is the covariant derivative. The constant
$\kappa$ is interpreted as the Hall conductivity.
The  second-order field equations are not integrable \cite{LVW}; they admit
integrable reductions, though.
The simplest of these  is when time-dependence is eliminated;  then,
for a judicious choice of the  self-interaction potential $V(\phi)$,
 the system admits
finite-energy vortex solutions \cite{JPD}.
Here we focus our attention to another, space-like reduction 
 \cite{AGJPS}.
 Assuming independence from one spacelike coordinate
and adding a suitable kinetic term 
yields in fact, after elimination of the gauge field
using its equation of motion,
\begin{equation}
{\cal L}=i\phi^\star\partial_t\phi
-\frac{1}{2}\big\vert(\partial_x-i\kappa^2\rho)
\phi\big\vert^2-V,
\label{redlag}
\end{equation}
where $\rho=\vert\phi\vert^2$ is the particle density.
This is the model proposed in Ref. \cite{BERA} to describe 
the edge states in the QHE.

The field equations associated to (\ref{redlag}) read
\begin{equation}
\begin{array}{c}
i\partial_t\phi=-\frac{1}{2}(\partial_x-i\kappa^2\rho)^2\phi
-{\kappa^2}j\,\phi+\displaystyle\frac{\partial V}{\partial \phi^\star},
\\[4pt]
j=\frac{1}{2i}
\big[\phi^\star(\partial_x-i\kappa^2\rho)\phi
-\phi(\partial_x+i\kappa^2\rho)\phi)^\star\big].
\end{array}
\label{modeqs}
\end{equation}

Then the particle density and the current satisfy the continuity 
equation $\partial_{t}\rho+\partial_{x}j=0$.
Let us first assume that $V=0$.
Now the non-local transformation \cite{MIPA}
\begin{equation}
\psi=\left(\exp[-i\kappa^2\int^x\!\rho(y)dy]\right)\phi
\label{nltraf}
\end{equation}
takes (\ref{modeqs}) into the modified non-linear Schr\"odinger
equation in which the density in
the non-linearity has been replaced by the current, 
\begin{equation}
\begin{array}{c}
i\partial_t\psi=-\frac{1}{2}\partial_x^2\psi-2\kappa^2j\,\psi,
\\  \\
j=\frac{1}{2i}\big[{\psi^\star\partial_x\psi
-\psi(\partial_x\psi)^\star\big].
}
\end{array}
\label{jnls}
\end{equation}

\section{The variable--coefficient NLS}

Decomposing $\psi$ into module and phase,
$\psi=\sqrt{\rho}\,e^{i\theta}$, yields (formally) the 
ordinary cubic NLS with variable coefficient,
\begin{equation}
i\partial_t\psi=-\frac{1}{2}\partial_x^2\psi
-F(t,x)\vert\psi\vert^2\psi,
\label{vnls}
\end{equation}
with 
$F(t,x)=2\kappa^2\partial_x\theta$.
Then Aglietti et al. \cite{AGJPS} observe that, for 
$
\theta=vx-\omega t,
$
Eq. (\ref{vnls}) reduces to the usual non-linear 
Schr\"odinger equation with
constant coefficient $F=2\kappa^2v$ which admits, for example,
 the travelling soliton solution
\begin{equation}
\psi_s=\pm e^{i(vx-\omega t)}\sqrt{\frac{1}{2\kappa^2 v}}
\frac{\alpha}{\cosh{\alpha(x-vt)}},
\qquad
\alpha^2=v^2-2\omega.
\label{trasol}
\end{equation}
The non-linearity in (\ref{vnls}) has to be attractive, $F>0$;
the solution (\ref{trasol}) is therefore chiral, $v>0$.

It is natural to ask whether the travelling soliton 
(\ref{trasol}) can be generalized. 
Let us first study the variable-coefficient NLS (\ref{vnls}) on its own.
It has been shown \cite{JOSHI} that this equation
only passes
the Painlev\'e test of Weiss, Tabor and Carnevale \cite{WTC},
when the coefficient of the non-linearity is 
\begin{equation}
F=(a+bt)^{-1},
\end{equation}
where $a$ and $b$ constants.
 For $b=0$,
$
F(t,x)
$
 is a constant and we recover the constant-coefficient NLS.
For $b\neq0$, the equation becomes explicitly time-dependent.
Assuming,
for simplicity, that $a=0$ and $b=1$, it reads
\begin{equation}
i\partial_{t}\psi+\frac{1}{2}\partial_{x}^2\psi
+{1\over t}\vert\psi\vert^2\psi=0.
\label{tnls}
\end{equation}

This equation can also be solved. Generalizing the usual
travelling soliton, 
 we find, for example, the $1$-soliton
\begin{equation}
\psi_0(t,x)=\frac{1}{\sqrt{t}}
\frac{e^{i(x^2/4t-1/2t)}}{\cosh\big[-x/t-x_{0}\big]}.
\label{travwave}
\end{equation}

 It is worth pointing out that the steps followed in
constructing (\ref{travwave}) are essentially the same as
those for the travelling soliton of the ordinary NLS ---
and this is not a pure coincidence.
A short calculation shows  in fact
that
\begin{equation}
\psi(t,x)={1\over\sqrt{t}}\exp\Big[{ix^2\over4t}\Big]\,
\Psi\big(-{1/t},-{x/t}\big)
\end{equation}
 satisfies the time-dependent equation (\ref{tnls}) 
if and only if $\Psi(t,x)$ solves Eqn. (\ref{vnls}) with $F=1$.
Our soliton (\ref{travwave}) comes in particular from the
 ``standing soliton'' $\Psi_{s}=\exp(it/2)\big(\cosh[x-x_{0}]\big)^{-1}$
 solution of the NLS.

\section{Non-relativistic conformal transformations}

Where does the formula (\ref{travwave}) come from~?
The non-linear space-time transformation
\begin{equation}
D~:
\left(\begin{array}{c}
t\\
x\\ 
\end{array}\right)
\to
\left(\begin{array}{c}
-\displaystyle{1/t}
\\  
-\displaystyle{x/t}
\\ 
\end{array}\right)
\label{dtraf}
\end{equation}
has already been met in a rather different context, namely
in describing planetary motion when the gravitational ``constant''
changes inversely with time, as suggested by Dirac \cite{DIR}.
One shows in fact that 
$
\vec{r}(t)=t\,\vec{r}^*\big(-{1/t})
$ 
describes  planetary 
motion with Newton's ``constant'' varying as 
$
G(t)={G_0/t},
$
whenever
$\vec{r}^*(t)$ describes ordinary planetary motion, i.e. the one
with a constant gravitational constant $G_0$ \cite{VIN}, \cite{DHP}.
The strange-looking transformation (\ref{dtraf})
 is indeed related to the
conformal structure of non-relativistic space-time \cite{JNH}.
It has been noticed in fact almost thirty years ago, that the 
``conformal''
space-time transformations
\begin{equation}
\begin{array}{ccc}
\left(\begin{array}{c}
t\\  
x\\ 
\end{array}\right)\hfill
\to
\left(\begin{array}{c}
T\\  
X\\ 
\end{array}\right)
=
\left(\begin{array}{c}
\displaystyle{\delta^2}
t\\ 
\displaystyle{\delta}\, 
x\\
\end{array}\right),
\qquad\hfill
&0\neq\delta\in{\bf R}\quad\hfill
&\hbox{dilatations}\hfill
\\ \\
\left(\begin{array}{c}
t\\  x\\ 
\end{array}\right)
\to
\left(\begin{array}{c}
T\\  
X\\
\end{array}\right)
=
\left(
\begin{array}{c}
\displaystyle\frac{t}{1-\kappa t}
\\ 
\displaystyle\frac{x}{-\kappa t}\\ 
\end{array}\right),
\qquad\hfill
&\kappa\in{\bf R}\hfill
&\hbox{expansions}\hfill
\\ \\
\left(\begin{array}{c}
t\\  
x\\
\end{array}\right)
\to
\left(\begin{array}{c}
T\\  
X\\
\end{array}\right)
=
\left(\begin{array}{c}
t+\epsilon\\  x\\ 
\end{array}\right),
\qquad\hfill
&\epsilon\in{\bf R}\hfill
&\hbox{time translations}\hfill
\\ 
\end{array}
\label{o21}
\end{equation}
implemented on wave functions according to
\begin{equation}
\Psi(T,X)=\left\{
\begin{array}{c}
\displaystyle{\delta}^{1/2}u(t,x)\hfill
\\
(1-\kappa t)^{1/2}\exp
\Big[i\displaystyle{\kappa x^2\over4(1-\kappa t)}\Big]\psi(t,x)
\hfill
\\
\psi(t,x)
\hfill
\\ 
\end{array}\right.
\label{impl}
\end{equation}
permute the solutions of
 the free Schr\"odinger equation. 
In other words, they are  symmetries of the free Schr\"odinger 
equation. The generators in (\ref{o21}) span in fact an ${\rm SL}(2,{\bf R})$ 
group. 
(A Dirac monopole, an Aharonov-Bohm 
vector potential and an inverse-square potential can also be included).
The  transformation $D$ in Eqn. (\ref{dtraf}) belongs to
this symmetry group: it is in fact 
 (i)  a time translation with 
$\epsilon=1$, (ii) followed by an expansion with 
$\kappa=1$, (iii) followed by a second time-translation with
$\epsilon=1$.
It is hence a symmetry for the free (linear) Schr\"odinger
equation.

The cubic NLS with constant non-linearity 
 is not more $SL(2,{\bf R})$ invariant: the 
 transformation $D$ in (\ref{dtraf}) implemented as in Eq. (\ref{impl})
 carries the cubic term into the
time-dependent term
$(1/t)\vert u\vert^2u$, just like Newton's gravitational potential
$G_{0}/r$ with $G_0={\rm const}$. is carried into the time-dependent
Dirac expression $t^{-1}G_{0}/r$.
More generally, the the non-linear Schr\"odinger equation
with time--dependent coefficients,
\begin{equation}
i\partial_t\psi+p(t)\partial_{x}^2\psi+F(t)\vert u\vert^2\psi=0,
\end{equation}
can be transformed into the constant--coefficient form, whenever
\begin{equation}
p(t)=F(t)\left(a+b\int^tp(s)ds\right)
\end{equation}
This same condition was found later as the one needed for
the Painlev\'e test \cite{JOSHI}.

Similar arguments explain the integrability of other NLS type
equations. For example, electromagnetic waves in a non-uniform
medium propagate according to
\begin{equation}
i\partial_{t}\psi+
\partial_{x}^2\psi+\big(-2\alpha x+2\vert\psi\vert^2\big)\psi=0,
\label{CLLeq}
\end{equation}
which can again be solved by inverse scattering \cite{CL}.
This is explained by observing that the potential term here
can be eliminated by  switching to a uniformly accelerated frame: 
\begin{equation}
\begin{array}{c} 
\psi(t,x)=\exp\big[-i(2\alpha xt+\frac{4}{3}\alpha^2 
t^3)\big]\Psi(T,X),\hfill
\\ \\
T=t,
\qquad
X=x+2\alpha t^2.\hfill
\\
\end{array}
\label{accelf}
\end{equation}
Then $u(t,x)$ solves (\ref{CLLeq}) whenever $\Psi(T,X)$ solves 
the free equation.

The transformation (\ref{accelf}) is again related to the structure of
non-relativistic
space-time. It can be shown in fact \cite{DHP} that the (linear) 
Schr\"odinger
equation
\begin{equation}
i\partial_{t}\psi+
\partial_{x}^2\psi-V(t,x)\psi=0
\end{equation}
can be brought into the free form
by a space-time transformation
if and only if the potential is
$
V(t,x)=\alpha(t)x\pm{\omega^2(t)\over4}x^2.
$
 For the uniform force field ($\omega=0$)
the required  transformation is precisely (\ref{accelf}).

For the oscillator potential ($\alpha=0$), one can use rather
Niederer's transformation~\cite{NI}
\goodbreak
\begin{equation}
\begin{array}{c}
\psi(t,x)=\big(\cos\omega t)^{-1/2}\,
\exp\big[-i{\omega\over4} x^2\tan\omega t\big]\,\Psi(T,X),\hfill
\\ \\ 
T=\displaystyle{\tan\omega t\over\omega}
\qquad
X=\displaystyle{x\over\cos\omega t}.\hfill
\\
\end{array}
\label{ntraf}
\end{equation}
Then 
\begin{eqnarray*}
i\partial_t\psi+\frac{1}{2}\partial_{x}^2\psi-{\omega^2x^2\over4}\psi
=
(\cos\omega t)^{-5/2}\,\exp\big[-i{\omega\over4}\tan\omega t\big]
\Big(i\partial_T\Psi+\frac{1}{2}\partial_{x}^2\Psi\Big)
\end{eqnarray*}
so that $\psi$ satisfies the oscillator-equations iff $\Psi$ solves the 
free equation. 

The Niederer transformation (\ref{ntraf}) leaves the inverse square
potential invariant; this explains why the Calogero model in a
harmonic background can be brought into
the pure Calogero form \cite{AVJE}.
Restoring the nonlinear term allows us to infer also
that
\begin{equation}
i\partial_t\psi+
\partial_{x}^2\psi+\Big(-{\omega^2x^2\over4}
+\frac{1}{\cos\omega t}\,
\vert \psi\vert^2\Big)\psi=0
\end{equation}
is integrable, and its  solutions are obtained from those of
the ``free'' NLS by the transformation (\ref{ntraf}).
Let us mention that the covariance w. r. t. 
chronoprojective transformations was used before
\cite{BUPE} for solving the NLS in oscillator and uniform--field backgrounds.

Now the constant-coefficient, damped, driven NLS,
\begin{equation}
i\partial_t\psi+
\partial_{x}^2\psi+F\vert\psi\vert^2\psi=
a(t,x)\psi+b(t,x),
\end{equation}
 passes the Painlev\'e test if
\begin{equation}
\begin{array}{c}
a(t,x)=\big(\frac{1}{2}\partial_t\beta-\beta^2\big)x^{2}+i\beta(t)
+\alpha_{1}(t)x+\alpha_{0}(t),\hfill
\\ \\
b(t,x)=0\hfill
\\
\end{array}
\end{equation}
\cite{CLA}, i. e., precisely when the potential can be
transformed away by our ``non-relativistic conformal 
transformations''.

\section{An integrable extension}

Unfortunately, the time-dependent travelling soliton (\ref{travwave})
is inconsistent with the original equation (\ref{jnls}), since its
phase is quadratic in $x$ rather than linear, as required by 
consistency. 
The clue for finding integrable extensions is to observe that
 Eqn. (\ref{jnls}) is in fact a Derivative Non-Linear Schr\"odinger
 equation (DNLS) \cite{MIPA}. Now the results of Clarkson and Cosgrove
 \cite{CLCO} say that the constant-coefficient equation
\begin{equation}
i\partial_t\psi+\frac{1}{2}\partial_{x}^2\psi
+ia\,\psi\psi^\star\partial_{x}\psi
+ib\,\psi^2\partial_{x}\psi^\star
+c\psi^3{\psi^\star}^2=0
\end{equation}
is integrable iff
\begin{equation}
c={1\over2}b(2b-a).
\label{clcocond}
\end{equation}

In our case $a=-b=-\kappa^2$ and $c=0$; 
Eq. (\ref{jnls}) is therefore not integrable. However, adding a
$6$th--order potential to the Lagrange density i. e.
considering rather
\begin{equation}
i\partial_{t}\psi+\frac{1}{2}\partial_x^2\psi
+2\kappa^{2}j\,\psi+{3\over2}\kappa^4\vert\psi\vert^4\psi
=0
\label{6jnls}
\end{equation}
converts (\ref{jnls}) into an integrable equation. 
Eq. (\ref{6jnls}) admits, e. g.,
the travelling wave solution
$\psi=\sqrt{\rho}\,e^{i\theta}$, where
\begin{equation}
\rho=\displaystyle\frac{\vert v\vert}{2\kappa^2}\,
\displaystyle\frac{1}
{\sqrt{2}\cosh\big[
v(x-\frac{v}{2}t)\big]+{\rm sign}\,v},\hfill
\qquad
\theta=\displaystyle\frac{v}{2}x.\hfill
\\ 
\label{newsol}
\end{equation}

This can be checked by observing that for
the Ansatz $\psi=f(x,t)e^{ivx/2}$ the modified NLS (\ref{6jnls}) 
again reduces to a constant-coefficient 
equation. Then the 
imaginary part of (\ref{6jnls}) requires
that $f(x,t)=f(x-(v/2)t)$, while the real part can be integrated
by the usual trick of multiplication by $f'$. 
The asymptotic conditions fix the integration
constant to vanish, yielding a six-order
non-linear equation, only containing
 even powers of $f$. Then, introducing $\rho=f^2$,
we end up with the equation
\begin{equation}
\big(\rho'\big)^2-v^2\rho^2+4\kappa^2 v\rho^{3}+4\kappa^4\rho^4=0,
\end{equation}
whose  integration provides us with 
(\ref{newsol}).

Another way of understanding how the integrability comes is to apply
 the non-local transformation
(\ref{nltraf}) backwards, which carries (\ref{6jnls})
into a Derivative Non-linear Schr\"odinger equation
of type II (DNLSII),
\begin{equation}
i\partial_t\phi+\frac{1}{2}\partial_x^2\phi
+2i\kappa^2\rho\,\partial_x\phi=0,
\end{equation}
which, consistently with Eq. (\ref{clcocond}),
 is integrable \cite{KNCLL}.
\smallskip

\goodbreak



\end{document}